# 4MOST Consortium Survey 6: Active Galactic Nuclei

Andrea Merloni[1]
David A. Alexander[2]
Manda Banerji[3]
Thomas Boller[1]
Johan Comparat[1]
Tom Dwelly[1]
Sotiria Fotopoulou[2]
Richard McMahon[3]
Kirpal Nandra[1]
Mara Salvato[1]
Scott Croom[4]
Alexis Finoguenov[1,5]
Mirko Krumpe[6]
Georg Lamer[6]
David Rosario[2]
Axel Schwope[6]
Tom Shanks[2]
Matthias Steinmetz[6]
Lutz Wisotzki[6]
Gabor Worseck[7]

[1] Max-Planck-Institut für extraterrestrische Physik, Garching, Germany
[2] Department of Physics, Durham University, UK
[3] Institute of Astronomy, University of Cambridge, UK
[4] Sydney Institute for Astronomy, University of Sydney, Australia
[5] University of Helsinki, Finland
[6] Leibniz-Institut für Astrophysik Potsdam (AIP), Germany
[7] Institut für Physik und Astronomie, Universität Potsdam, Germany

X-ray and mid-infrared emission are signposts of the accretion of matter onto the supermassive black holes that reside at the centres of most galaxies. As a major step towards understanding accreting supermassive black holes and their role in the evolution of galaxies, we will use the 4MOST multi-object spectrograph to provide a highly complete census of active galactic nuclei over a large fraction of the extragalactic sky observed in X-rays by eROSITA that is visible to 4MOST. We will systematically follow up all eROSITA point-like extragalactic X-ray sources (mostly active galactic nuclei), and complement them with a heavily obscured active galactic nuclei selection approach using mid-infrared data from the Wide-field Infrared Survey Explorer (WISE). The X-ray and mid-infrared flux limits of eROSITA and WISE are well matched to the spectroscopic capabilities of a 4-metre-class telescope, allowing us to reach completeness levels of ~80–90% for all X-ray selected active galactic nuclei with fluxes $f_{0.5-2\,\text{keV}} > 10^{-14}$ erg s$^{-1}$ cm$^{-2}$; this is about a factor of 30 deeper than the ROSAT all-sky survey. With these data we will determine the physical properties (redshift, luminosity, line emission strength, masses, etc.) of up to one million supermassive black holes, constrain their cosmic evolution and clustering properties, and explore the connection between active galactic nuclei and large-scale structure over redshifts $0 \lesssim z \lesssim 6$.

## Scientific context

The presence of a supermassive black hole (SMBH) at the centre of virtually every massive galaxy in the nearby Universe is a robust observational fact. However, their formation, growth, and connection to the evolution of galaxies and large-scale structure remain largely a mystery (for example, Alexander & Hickox, 2012). There is strong indirect evidence that the formation and growth of SMBHs and their host galaxies are closely related, but it is unclear what physical mechanisms are responsible for this close coupling. In order to discriminate between the varieties of different model predictions proposed in recent years, sizeable samples of active galactic nuclei (AGN) over different periods and phases in their evolution are needed. To this end, X-ray and mid-infrared (mid-IR) AGN searches are less biased by obscuration effects than optical ones and provide a solid foundation for the most comprehensive SMBH evolutionary studies. X-ray selected AGN are also more complete at the low end of the AGN luminosity function, where optical and mid-IR selection approaches are more affected by host galaxy light dilution.

Unfortunately, the current samples of X-ray and mid-IR selected AGN with spectroscopic redshifts are relatively small (a few thousand; see Figure 2) compared to the optically selected AGN samples available from large-area surveys like the Sloan Digital Sky Survey (SDSS). Consequently, the modest source statistics of X-ray and mid-IR selected AGN hamper our understanding of how the relationship between AGN and their host galaxies evolves as a function of redshift, AGN luminosity, nuclear obscuration, host galaxy mass, star formation properties, and large-scale environment. A complete AGN sample of hundreds of thousands of sources with spectroscopic redshifts and classifications is required to fully sample this multi-dimensional parameter space. The synergy between eROSITA (Merloni et al., 2012), complemented by VISTA near-IR and WISE mid-IR AGN selection, and 4MOST will allow this to become a reality.

The 4MOST AGN survey will provide spectroscopic identification for up to one million AGN out to redshifts of $z \sim 6$ over an area of about 10 000 square degrees (see Figure 1). The well understood X-ray and mid-IR AGN identification approaches, combined with the uniform and well characterised selection functions of the eROSITA and WISE all-sky surveys, will ensure a highly complete AGN selection, largely independent of the presence of nuclear obscuration. We will aim for a high level of spectroscopic completeness[a] for the 4MOST AGN survey to keep statistical uncertainties to a minimal level. The WISE-selected AGN will include a sample of luminous dust-obscured quasars, which may be responsible for powerful AGN-driven feedback (for example, Banerji et al., 2012) and may escape detection by eROSITA given its predominantly soft X-ray response. Such a complete 4MOST AGN sample will serve as a benchmark against which to test competing models of SMBH formation and growth, and their connection to galaxies and large-scale structure in the Universe.

## Specific scientific goals

Our science goals are set out below.

### Evolution of the most massive and powerful SMBHs

The X-ray luminosity function (XLF) of moderate- and low-luminosity AGN ($L_X \lesssim 10^{44}$ erg s$^{-1}$) is relatively well constrained as a result of extensive studies following up on deep Chandra and XMM-Newton small- and medium-area surveys (see Figure 2 and, for example, Aird et al., 2015). In contrast, the eROSITA- (and WISE-) selected samples in the 4MOST AGN survey will be geared towards the most luminous objects at any redshift, and are expected to provide an improvement of many orders of magnitude in the number of sources emitting above the break in the luminosity function (Figure 2). Redshift determination for these AGN will provide an unprecedented look at the evolution of the most massive and powerful SMBHs. Finally, in combination with optical and near-IR imaging surveys over the 4MOST survey footprints, we will also search for high-redshift quasar candidates at $z > 4$ to complement the tens of $z > 6$ X-ray selected AGN expected from eROSITA (see Figure 2).

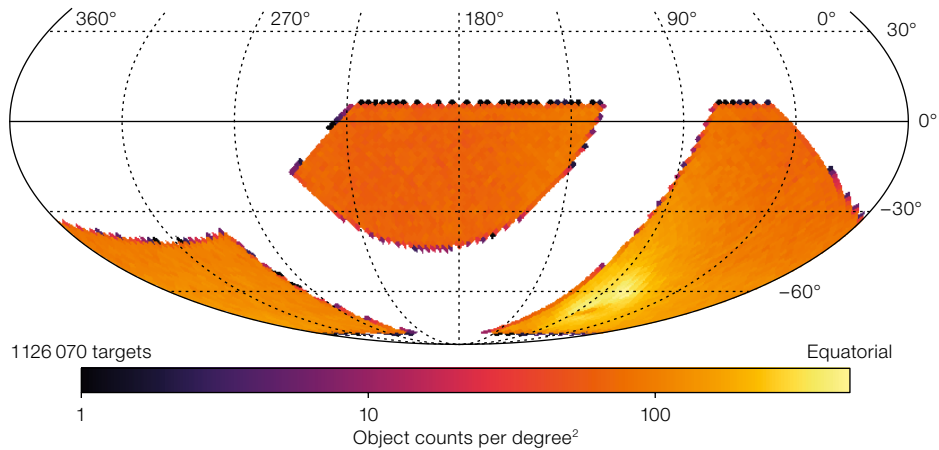

Figure 1. Sky density in equatorial coordinates of the 4MOST AGN targets, including all sub-survey components (based on the mock eROSITA catalogue of Comparat et al. [in preparation], and the current WISE-selected AGN catalogue).

### Frequency of accreting SMBHs in the galaxy population

The highly complete X-ray and mid-IR sample of AGN will provide a legacy dataset from which to measure the incidence of accreting SMBHs within: (1) the $z < 1$ galaxy population; (2) merging galaxies and other morphological classes (delivered by high-quality optical/near-IR imaging from, for example, surveys from the Hyper Suprime-Cam [HSC], the Large Synoptic Survey Telescope [LSST], and the Euclid mission); (3) radio galaxies (and jet-dominated AGN of various classes, as delivered by future deep and wide-area radio surveys); and (4) large-scale structures such as voids, filaments, groups, clusters (from synergy with SZ surveys, CMB lensing, and X-ray cluster surveys, including eROSITA itself). Given the excellent source statistics, the 4MOST AGN survey will also allow for the most comprehensive measurements to date of the accretion rate distribution of AGN as a function of redshift, luminosity, and host-galaxy properties.

Figure 2. The central panel shows the predicted X-ray luminosity-redshift plane spanned by AGN selected by eROSITA within the 4MOST X-ray Wide and Deep areas (orange/red contours), compared to all known X-ray sources with spectroscopic redshifts detected in several deep and wide survey fields (COSMOS, Lockman Hole, CDFS and CDFN, XMM-XXL, black dots; SDSS/SPIDERS DR14, grey dots; Dwelly et al., 2017). The thick green line is the approximate location of the break in the AGN X-ray luminosity function $L_X$ where the bulk of the accretion power is released. The thick cyan line is the approximate limit above which the IR Wide targets will be selected, assuming a standard infrared-to-X-ray conversion. The top histograms show the redshift distribution of eROSITA+4MOST sources (red/orange) compared to all other known X-ray sources (black, grey), while the right-hand histograms compare the respective luminosity distributions.

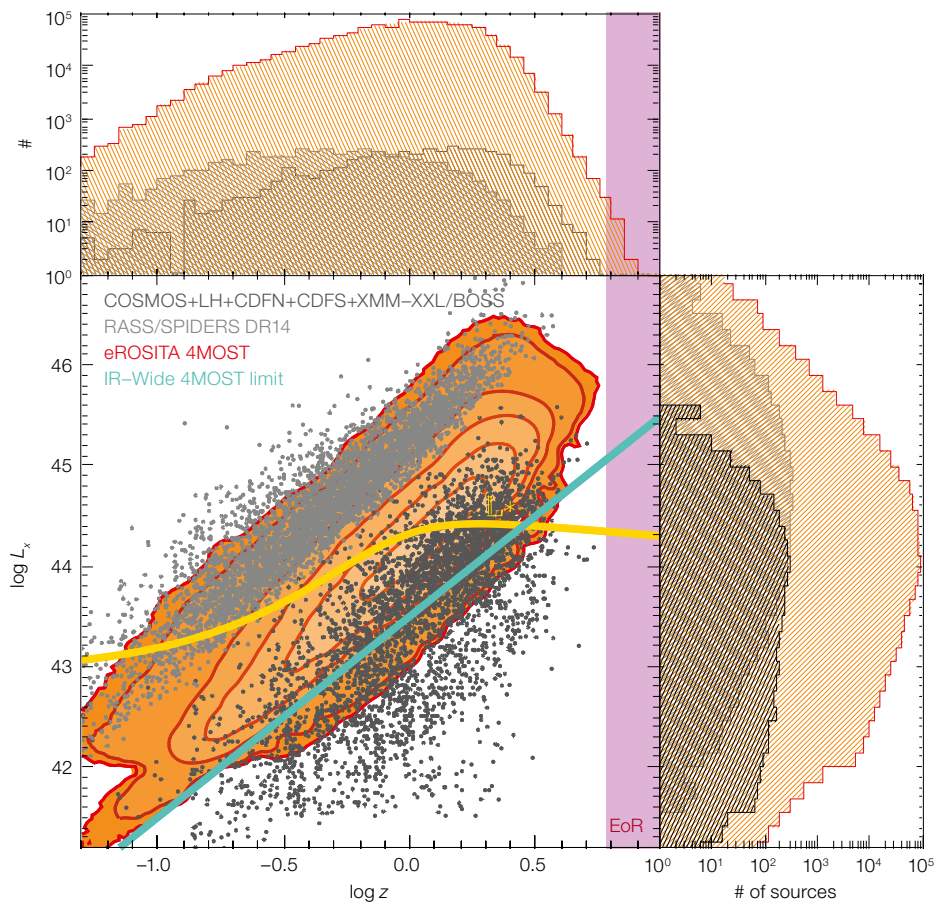

### SMBH and host-galaxy spectral measurements

The 4MOST optical spectra will allow for direct estimates of SMBH masses for the large subset of AGN with broad emission lines using virial relations (Shen, 2013). These measurements will probe the physics of accretion onto SMBHs, and will also allow for a direct investigation of the evolution of the SMBH-host galaxy scaling relations out to $z \sim 3$. For the population of lower-redshift AGN ($z < 1$) the well calibrated (see Science below) medium-resolution ($R \sim 6500$) spectra will also deliver quantitative measurements of a wide variety of AGN host-galaxy properties (stellar masses, star formation rates, dust reddening, stellar population ages, and metallicities; see, for example, Menzel et al., 2016) and the identification of AGN-driven outflows to constrain their impact on the star formation in the host galaxy (for example, Harrison, 2017). In addition, the LRS resolution will allow unambiguous identification of the [O II] doublet in the redshift range where it is the only feature in the spectrum with high signal to noise (S/N). This could be particularly important for the faint, obscured WISE AGN targets.

### AGN activity and the large-scale environment

The measurement of the spatial clustering of AGN has emerged as a major way to investigate the relationship between AGN, their host galaxies and the larger-scale environment. AGN clustering measurements have the potential to: (1) determine the distribution of AGN as a function of dark matter halo mass (Krumpe et al., 2015); (2) infer AGN triggering mechanisms and lifetimes; and (3) measure the cosmological parameters (for example, via baryon acoustic oscillation [BAO] measurements; Kolodzig et al., 2013; Comparat et al., 2019). Direct measurements of the 3D spatial clustering in X-ray selected AGN samples at $z > 1$ have mostly been limited to medium-deep area surveys of only a few square degrees. These studies suggest that $z \sim 1$ X-ray selected AGN are more clustered than optically selected AGN. However, the results remain uncertain owing to the effects of cosmic variance, small number statistics, and differences in luminosities. The large area and the sample size of eROSITA and WISE will significantly improve these studies, boosting our knowledge of the AGN clustering properties.

### Fundamental cosmological constraints

In recent years it has been demonstrated beyond doubt that the X-ray:ultraviolet luminosity ratios of unobscured quasars follow a tight non-linear relationship. This allows them to be used as "standard candles" to probe the geometry of the expanding universe out to $z \sim 6$, in much the same way that type Ia Supernovae have been used at $z < 1$ (Risaliti & Lusso, 2018). The large AGN sample from eROSITA will provide excellent source statistics to improve the current cosmological constraints, particularly thanks to a few $10^4$ $z > 3$ quasars, which offer unique probes of the expansion history of the universe at high redshift.

### Constraining AGN accretion physics via variability study

The survey area close to the South Ecliptic Pole (Deep survey, see below) will be revisited at X-ray wavelengths several times thanks to the eROSITA scanning strategy (Merloni et al., 2012), and will also be covered by a denser tiling of 4MOST pointings, because of the high source density in this region (which includes the LMC targets). This will allow for additional unique scientific opportunities for the study of time-variable AGN, including eclipsing sources, variable absorption and dramatic flux variability events ("changing look" AGN).

## Science requirements

The aforementioned scientific goals require a highly complete and uniform set of spectroscopic redshift measurements over a very large area. The uniformity of the selection, necessary to achieve most of the science goals, will be guaranteed by the well-understood and uniform eROSITA and WISE selection functions, provided that nearly complete counterpart identification and successful spectroscopic redshift measurements are achieved. These simple goals drive most of the science requirements, which we list here.

– To cover an overall area of ~ 10 000 square degrees.
– To target both the point-like and the extended optical source counterparts of eROSITA- and WISE-selected AGN with an average sky density of ~ 100–120 per square degree.
– To observe the area around the South Ecliptic pole (~ 300 square degrees) more frequently (with a cadence of at least half a year) and deeper with a target density of ~ 200 per square degree.
– To robustly measure redshifts for faint targets (down to $r = 22.5$–$23.0$ magnitudes) and to derive physical characteristics of the brighter sources from high S/N optical spectra.
– To achieve good (better than 10% accuracy) relative flux calibration so as to use AGN and host galaxy continuum measurements to derive physical characteristics from continuum and emission-line flux measurements.

## Target selection and survey area

The goal for the 4MOST AGN survey is to obtain spectra for ~ 80–90% of the X-ray and mid-IR selected AGN samples to yield redshifts for up to one million AGN. The survey area is defined by the eROSITA all-sky X-ray survey whose proprietary data rights lie with the MPE-led German Consortium. eROSITA will observe most of the area to a relatively uniform depth with two deeper regions around the ecliptic poles. We have therefore defined two separate sub-components to the survey: the X-ray Wide Survey, covering almost all of the German eROSITA extragalactic sky visible with 4MOST (10 000 square degrees); and the X-ray Deep Survey, covering a circle of ~ 300 square degrees centred on the South Ecliptic Pole. To these will be added a third component, the IR WIDE Survey, which shares the footprints with the X-ray Wide area, and selects targets based on their WISE mid-IR properties.

The target selection of the eROSITA sources is straightforward — all eROSITA X-ray detected point sources within the 4MOST extragalactic footprints will be targeted with the Low-Resolution Spectrograph (LRS) fibres. The eROSITA PSF (~ 28-arcsecond half energy width) is sufficient to distinguish X-ray AGN (and stars) from the most common extended X-ray sources (clusters of galaxies; Clerc et al., 2018). In the X-ray Wide Survey, we expect ~ 90% of the eROSITA sources with $f_{0.5-2\,keV} > 10^{-14}$ erg s$^{-1}$ cm$^{-2}$ to be brighter than $r = 22.8$ magnitudes; in the X-ray Deep Survey we expect ~ 90% of the sources with $f_{0.5-2\,keV} > 5 \times 10^{-15}$ erg s$^{-1}$ cm$^{-2}$ to be brighter than $r = 23.4$ magnitudes, and this sets our expected survey depth. These X-ray selected AGN will be complemented with near- and mid-IR AGN selected from the VISTA Ks and WISE W1-W3 bands using colour selection techniques (for example, Mateos et al., 2012), potentially supplemented with a machine-learning approach to improve the AGN-selection efficiency and completeness (for example, Fotopoulou & Paltani, 2018). The eROSITA X-ray sources will provide the majority of the AGN sample, but the WISE mid-IR AGN will be important to select the heavily obscured AGN, which will be missed because of the relatively soft X-ray response of eROSITA, to give a highly complete 4MOST AGN sample.

| Name | Area (deg$^2$) | Average density (deg$^{-2}$) | Limiting magnitude | Number of sources (10$^3$) | Selection notes |
|---|---|---|---|---|---|
| X-ray Wide | 10 000 | ~ 90 | $r < 22.8$ | ~ 800 | $F_{0.5-2\,keV} > 10^{-14}$ erg s$^{-1}$ cm$^{-2}$ |
| IR Wide | 10 000 | ~ 20 | $r < 22.8$ | ~ 180 | See Mateos et al., 2012 |
| X-ray Deep | 300 | ~ 200 | $r < 23.2$ | ~ 50 | $F_{0.5-2\,keV} > 5 \times 10^{-15}$ erg s$^{-1}$ cm$^{-2}$ |
| High-$z$ Quasars | 10 000 | ~ 4 | $i_{AB} < 22.5$ ($z > 5$) $z_{AB} < 22.0$ ($z > 6$) | ~ 40 | See Reed et al., 2017 |

Table 1. Summary of 4MOST AGN survey characteristics: (1) area covered, requirements; (2) average target density; for the IR Wide survey, we quote the total number of unique targets, after accounting for duplicates with the X-ray Wide survey; (3) approximate limiting optical magnitudes; (4) total number of spectra to be obtained in order to satisfy the spectral success criteria.

Finally we will also carry out a High-$z$ Quasar Survey using a combination of optical, near-IR and mid-IR data (for example, from DES, HSC, LSST, VISTA and WISE) to $i < 22.5$ magnitudes for $5 < z < 6$ selection, $z < 22.0$ magnitudes for $6 < z < 6.5$, and spectral energy distribution based techniques to select and classify the targets (for example, Reed et al., 2017).

## Spectral success criteria and figure of merit

The spectral success criteria are defined using an empirical relation between the redshift measurement success and S/N over different broad bands from the Baryon Oscillation Spectroscopic Survey (BOSS) of X-ray selected AGN (Menzel et al., 2016). The BOSS campaign targeted XMM-Newton AGN with counterparts brighter than $r = 22.5$ magnitudes to be representative of the 4MOST targets. Taking a < 3% failure rate as a target threshold, and scaling for the different resolutions of the BOSS and 4MOST LRS, we obtain the following S/N estimates per pixel thresholds: 2.1, 2.4 and 2.8 for the three arms of the 4MOST spectrograph (i.e., blue, green and red). It is expected that the above criteria are very conservative, as they are based on continuum measurements for a set of objects which should contain numerous narrow emission/absorption lines, which will be used to successfully measure redshifts. We envisage that all counterparts of eROSITA-detected point-like X-ray sources will be observed by 4MOST with up to a maximum of ~ 2 hours exposure. The spectral success criteria for the WISE-selected AGN will be comparable to the eROSITA-selected AGN, given the similar distribution of optical magnitudes and redshifts (for example, Lam, Wright, & Malkan, 2018).

The overall survey figure of merit definition is driven by the main requirement to reach the highest possible level of spectroscopic completeness (minimum requirement of 90% for the wide surveys, and 80% for the deep one) over an area of 10 000 square degrees (Wide) and 300 square degrees (Deep).

### Notes

[a] Here we define completeness as the fraction of X-ray and mid-IR sources in the targeting samples for which we successfully acquire spectroscopic redshift measurements.